# Comment on "Comparing gold nano-particle enhanced radiotherapy with protons, megavoltage photons and kilovoltage photons: A Monte Carlo simulation" by Lin *et al* [Phys. Med. Biol. 59 (2014) 7675–7689]

Hans Rabus [1]

[1] Physikalisch-Technische Bundesanstalt (PTB), Berlin, Germany

E-mail: hans.rabus@ptb.de

**Abstract**

In their article published in Phys. Med. Biol. 59 (2014) 7675–7689, Lin *et al* studied the dose enhancement of gold nanoparticles (GNPs) for proton therapy, which they compared with the case of photon irradiation. This comment points out two caveats to the methodology used by Lin *et al* that may not be evident to readers and may contribute to confusion in the literature about the dose enhancement by gold nanoparticles.

---

## 1. Assessment of the dose enhancement by gold nanoparticles

In the paper of Lin *et al* (2014), dose enhancement by gold nanoparticles (GNPs) is assessed by Monte Carlo simulations of radiation transport in a three-step procedure. First, simulations of radiation transport in water were performed to determine the fluence of ionizing particles impinging on a GNP. Then, the fluence of electrons produced by the radiation impinging on a spherical GNP with diameter of 50 nm or a water volume of the same size (a 'water nanoparticle (WNP)') was determined.

In the third step, the energy imparted by the electrons leaving the GNP or the WNP was scored for volumes at different distances from the GNP surface. The ratio of the energy imparted to the mass of these volumes was then interpreted as the "dose", and the dose enhancement factor (DEF) was defined as the ratio between the "dose" obtained with the GNP and the "dose" obtained with the "WNP".

In reality, the quantity determined in step 3 of the simulations is the dose *contribution* of the electrons emitted from the GNP or the "WNP" after the radiation interaction in the GNP or "WNP". Therefore, defining the ratio of these two contributions as the DEF is extremely misleading and gives values that do not correspond to dose enhancement.

Before detailing the arguments already presented in (Rabus *et al.* 2019, 2020, 2021), it may be useful to examine a simple example. Consider a target volume of the same size as the GNP, with its center located at a radial distance of 250 nm from the GNP center. Within 250 nm of the center of this target volume, there is a GNP of 50 nm diameter and about 399 "WNPs" of 50 nm diameter. If the GNP is not present, there are 400 "WNPs." Thus, considering only the dose contribution (DC) of nanoparticles at 250 nm from the target volume, this dose contribution is that of one GNP and 399 "WNPs" when the GNP is present, and that of 400 "WNPs" when the GNP is absent. The ratio of these two dose contributions, the dose contribution enhancement factor (DCEF) of sources at 250 nm from the target volume is

$$DCEF(250\ nm) = 1+(DC(GNP)/DC("WNP")-1)/400 \tag{1}$$

Thus, even if the DC of a GNP for protons is a factor of 5 to 15 greater than the DC of a "WNP" (Fig. 3(c) in (Lin *et al* 2014)), the ratio DCEF(250 nm) is only about 1.025. And even if for photons the DC of a GNP is a factor of 1000 greater than that of a "WNP", the DCEF(250 nm) drops to about 3.5.

However, in addition to the contribution from the GNP and "WNPs" at 250 nm, there are also contributions from "WNPs" at smaller and larger distances from the target volume. In the absence of the GNP, these dose contributions simply add up to the prescribed dose $D_w$. With the GNP present, the absorbed dose $D_{t,g}$ in the target volume is approximately given by

$$D_{t,g} = DC_{GNP}(|\vec{r}_t|) + \frac{1}{V_{NP}} \int_{V'} DC_{WNP}(|\vec{r} - \vec{r}_t|)dV = DC_{GNP}(|\vec{r}_t|) - DC_{WNP}(|\vec{r}_t|) + D_w \tag{2}$$

where $V_{NP}$ is the nanoparticle volume, $DC_{WNP}$ is the dose contribution of a "WNP" located at $\vec{r}$, $\vec{r}_t$ is the target position, $DC_{GNP}$ is the dose contribution from the GNP (located at the origin), and the integration domain $V'$ is defined by the range of the secondary electrons and excludes the volume of the GNP. Therefore, the dose enhancement factor is given by

$$DEF(|\vec{r}_t|) = 1 + [DC_{GNP,p}(|\vec{r}_t|) - DC_{WNP,p}(|\vec{r}_t|)]/D_{w,p} \tag{3}$$

where $DC_{GNP,p}$ and $DC_{WNP,p}$ are the dose contributions from a GNP and a "WNP" per primary particle (shown in Fig. 3 and 4 (a) and (b) of (Lin *et al* 2014)) and $D_{w,p}$ is the dose to water for a primary particle fluence $\Phi_p$ of 1 per GNP cross-section, i.e., $\Phi_p = 5.1 \times 10^{10}$ cm$^{-2}$. $D_{w,p}$ can be estimated for protons from the stopping power (Berger *et al* 2005) and for photons from the the mass-energy absorption coefficient (Hubbell and Seltzer 2004). For the monoenergetic 50 keV photon irradiation, this gives 0.016 Gy. From Fig. 4 (a), this is approximately the dose contribution from the GNP at about 20 nm from the GNP surface, where then the DEF is 2 instead of $10^3$. At $10^3$ nm from the GNP surface, where the "DEF" increases in Fig. 4 (c) of Lin *et al* (2014), the true DEF is about 1.0007 according to eq. 3.

For 10 MeV and 100 MeV protons, values of $D_{w,p}$ = 370 Gy and $D_{w,p}$ = 59 Gy, respectively, are obtained. The dose contribution per incident proton at the GNP surface is about 25 Gy and 2.5 Gy at these two proton energies, and the resulting DEF values at the surface of the GNP are 1.07 and 1.04 respectively. At 1000 nm from the GNP, where the "DEF" in Fig. 3(c) of Lin *et al* (2014) saturates at about 15, the actual DEF for both energies is below 1.0001. (These values were obtained using eq. 3 based on estimated dose values from the figures in Lin *et al* (2014)

## 2. Particle fluence estimations

In Sections 2.2.3 and 2.2.4 of their paper, Lin *et al* (2014) describe briefly how they proceeded to estimate the fluence of particles interacting with the GNPs when the primary particles have an energy spectrum. Briefly, they performed radiation transport simulations in a water phantom and scored all particles traversing a circular area with a radius of 25 mm to obtain a distribution of phase-space coordinates. This distribution was then modified so that the lateral coordinates fell within a circle with a radius of 25 nm, and the direction of motion was changed so that the particles moved along the direction of the incident beam. A $1/\cos\theta$ correction was applied for the weight of the particles. Apart from this description of the procedure, no justification for its correctness is given.

The background for the method used by Lin *et al* (2014) is the so-called common random number (CRN) approach. Let us assume that the primary particles are emitted from a plane along the normal direction and let us refer to the collection of particles and energy transfer points that occur during the simulation of a primary particle a "shower". This shower is characterized by a sequence of random numbers used in modeling of the various interactions that occur. If the emission point of the primary particle is shifted in the source plane, and if the same sequence of random numbers is used in the simulation, then all energy transfer points of the shower will be shifted by the same vector in a direction parallel to the source plane.

Thus, assume a particle resulting from a primary particle starting at a given position in the source plane traverses the plane used for scoring at a location and along a direction of motion that does not transport the particle to the GNP. Applying a lateral shift so that the trajectory of the particle from the shifted position on the scoring plane hits the GNP corresponds to a position on the source plane shifted laterally by the same vector. If the source emits primary particles uniformly, all starting positions are equivalent. Therefore, the lateral displacement simply "picks" a position on the source plane that is as likely as the one used in the simulation but with the benefit that the trajectory intersects the GNP. The factor $1/\cos\theta$ accounts for the fact that the area covered by possible starting points on the source plane is generally not a circle (like the cross-section of the GNP), but an ellipse with one axis elongated by this factor $1/\cos\theta$.

Estimation of the particle fluence at the GNP using this approach is justified when the possible lateral offsets from the starting position of the primary particle is small compared to the lateral extent of the area used for the scoring. Under these conditions, the occurrence of starting positions outside the part of the surface plane from which primary particles are emitted is rare and negligible. For protons, this should be the case. For photons, on the other hand, using the CRN approach would actually require an infinite lateral extension of the primary radiation field. Practically, this could be achieved by choosing the size of the beam much larger than the area used for scoring. In addition, the radiation field for photons also depends on the extent of the geometry in the direction of the primary particle beam. As shown by Rabus *et al* (2019) for a geometry with infinite depth and a beam of infinite width, the contribution of scattered photons is by a factor of order 5 higher than the fluence of primary beam for kilovoltage X-ray and by orders of magnitude for Co-60 radiation.

In addition, it should be noted that both Lin et al (2014) and Rabus et al (2019) estimated the fluence of particles at the GNP using a simulation of radiation transport in water. This is appropriate when the case of a single GNP is considered. If the concentration of gold atoms in the volume loaded with GNPs becomes significant, this will also change the particle fluence spectra within this volume. In principle, this can be remedied by performing the simulation of the first step not for pure water, but for a uniform mixture of gold and water.

## Conclusions

From the above arguments, it appears that for photon fields whose fluence is determined by the procedure used by Lin *et al* (2014), the fluence of particles impinging on the GNP may be underestimated, while the procedure is expected to be correct for protons. With respect to determining the dose enhancement around a single GNP, the approach of Lin *et al* (2014) to use the ratio of the dose contribution by electrons produced in the GNP to that of electrons produced in a water volume with the same dimensions resulted in an overestimation of the DEF by orders of magnitude.

## Supplement

This Comment was submitted (without this supplement) to Physics in Medicine and Biology in December 2022. It was reviewed by members of the editorial board, who recommended that the comment be rejected. Because I believe that the arguments made by the reviewers indicate that they felt the approach taken in writing this comment was disrespectful to the authors or even that they themselves were offended and suspected malicious intent on my part, I would like to add the following statements here for clarification:

- ***The intent of this comment is not to blame the authors of the article or the peer-review process for this article.***
- ***In my view, science is a collective learning effort and peer review is the best approach we have to assure quality in this process.***

I assume that the members of the editorial board were driven by a - very commendable - desire to protect their authors from disrespectful attacks, and probably had serious concerns about a mudslinging match might break out between the commentator and the authors. The former was not intended, and the latter is not imminent, as I pointed out in my appeal against the first rejection notice. I therefore personally feel that the arguments on which the first and the final decision were based were disrespectful to me. The interested reader may refer to the attached reproduction of the email correspondence. (For clarification, the decision on this comment was communicated to me only after I appealed the decision to reject a second comment on another paper.)

**Your manuscript PMB-114492 - Decision on your manuscript**
Physics in Medicine and Biology
An hans.rabus
Kopie hans.rabus

23.01.2023 18

Details verberg

Von "Physics in Medicine and Biology" <onbehalfof@manuscriptcentral.com>
An hans.rabus@ptb.de
Kopie hans.rabus@ptb.de
Protokoll: Bitte Antwort an pmb@ioppublishing.org
Diese Nachricht wurde beantwortet.

Dear Dr Rabus,

REVIEWER REPORT(S):

Referee: 1

COMMENTS TO THE AUTHOR(S)
The submission of comments on a several-year old paper may not be timely enough. I would recommend the author to send a personal letter to the authors of the objected paper to give them a chance to handle the problems or to write his own paper to draw new valid conclusions.

Referee: 2

COMMENTS TO THE AUTHOR(S)
[editorial board member report]

I have been asked to assess the submitted Comment below, as well as another Comment by the same author for another paper by Lin et al.

I have concerns with this approach to communicating disagreements with published work. The Comments focus on differences of opinion regarding clinical feasibility and read like a referee report ought to. The Commenting author may not approve of the assumptions made or the specific approaches of Lin et al., but if there are no major errors, I don't see the need for publication of the Comment. I would suggest that the Commenter should write their own original article approaching the same topic, but it appears that they already have.

The purpose of the Comments seem to be to alert readers that the published work of Lin et al. used unrealistic assumptions, and therefore the results should not be trusted. However, in my opinion, Monte Carlo simulations of nanoparticle enhancement all suffer from some sort of unrealistic assumptions and still generally underestimate the biological effects of nanoparticles in radiation therapy anyways. Arguing over these minor points rather than addressing the bigger issues facing this therapy does not seem like the best use of time or resources.

I also wonder whether Lin et al. will be given an opportunity to respond to the criticisms in the Comments and what the mechanics of that will be. Ultimately, this could be an interesting discussion, but I don't think that it is appropriate to have that play out over months/years through this Comment mechanism. What I would like to see is for the Comments' author and Lin et al. to collaborate offline to produce a follow up article addressing the issues together.

Ultimately, the question I come back to is: "how will these Comments help patients with cancer?" I don't really see that they do, and therefore I don't see a need for publication.

Letter reference: DEC:RejBM:S

**Antwort: Your manuscript PMB-114492 - Decision on your manuscript**
Hans Rabus an pmb

27.01.2023 18:2
Details verberge

| Von | Hans Rabus/PTB |
| --- | --- |
| An | pmb@ioppublishing.org |

| | Physics in Medicine Your manuscript PMB-114492 - Decision on your manuscript | 23 Jan 18:34 | 11K |
| --- | --- | --- | --- |
| | Hans Rabus — *Dear Editor, thank you for your email informing me of your decision on my submitted Comment. Given* | | |
| | Hans Rabus — *Dear colleagues, since it appears that I did not yet get a reply nor a confirmation of receipt, I am* | | |

Dear Editor,

thank you for your email informing me of your decision on my submitted Comment.

Given that my motivation for writing the Comments is fully aligned with some of the considerations that make Reviewer 2 recommend rejection, I make another attempt of rebuttal.

Reviewer 2 is concerned that publishing the Comment could be the start of a futile debate that might be a waste of resources that would be of no use for patients with cancer.

The fear that a feud would be started is completely unsubstantiated. The referees' reactions to the Comment are in clear contrast to those of the two co-authors of the papers, who supervised the lead author when she was working at MGH before quitting scientific research. Harald Paganetti's replied that "Such discussions are important" and Jan Schuemann even offered me to become a co-author of the Comment to give it more weight. Although such an approach would have avoided any potential concerns about malevolent intentions on my side, most of the reviewers' arguments could then have been raised as well. A confirmation of the caveats raised by the Comment in a Reply by the authors would give them even more weight in my view. In any case, I am confident that the discussion will be brief, factual and very constructive.

The main motivation for me to write the Comments in the first place was that resources are wasted when other authors copy approaches and assumptions of papers like those of Lin et al. without being aware of the caveats. "The Comments read like a referee report should" because they are extracts of a detailed referee report I have recently written for another journal in the area of medical physics, where I reviewed a manuscript that was largely inspired by the papers of Lin. et al. 2014, 2015 in PMB and essentially copied their methodology and choice of parameters of the simulations.

I fully agree with the reviewer that many published Monte Carlo simulations of nanoparticle enhancement generally underestimate the biological effects nanoparticles may have in radiation therapy. At least, when gold nanoparticle concentrations are considered that have actually been reported so far in preclinical or radiobiological studies. (There are exceptions such as McMahon et al., Radiother Oncol 100 (2011) 412–416.)

I disagree that this failure of Monte Carlo simulations is due to unrealistic simplifications. From my experience as reviewer and supervisor of students, the vast majority of these studies suffer from either considering extreme cases where giant effects were found, such as with the papers of Lin et al., or methodological deficiencies that end up comparing apples with oranges. In consequence, conflicting conclusions can be found in literature on questions such as optimum nanoparticle size and colleagues trying to review the field (e.g., Moradi et al., RPC 180 (2021) 109294) have to limit their endeavors to simply reporting "A did M and obtained X while B did N and obtained Y".

For a patient suffering from cancer today it does not make a difference whether the Comments are published or not. But this "no immediate benefit" catch may apply to many papers published in PMB. Suppressing the Comments means continuing the waste of resources and efforts of early-stage researchers or other newcomers to the field who copy the methodology of published work not knowing that there are caveats and plausibility concerns. This will perpetuate the confusion in literature and hamper progress in our understanding of nanoparticle radiation effects and delay the full exploitation of this promising therapy.

Starting the discussion now gives us the possibility to change this situation, make the field focus on the real challenges and eventually make the introduction of nanoparticles in clinical practice happen earlier and have future patients suffering from cancer benefit from this modality. Suppressing the scientific discussion will block this way.

It is up to the Editorial board to decide which alternative PMB should support.

With kind regards

Hans Rabus

## A message about manuscripts PMB-114492 & PMB-114487
Physics in Medicine and Biology
An hans.rabus

10.02.2023 13:04
Details verbergen

Von "Physics in Medicine and Biology" <onbehalfof@manuscriptcentral.com>
An hans.rabus@ptb.de
Bitte Antwort an pmb@ioppublishing.org

Dear Dr Rabus,

Re: "Comment on "Comparing gold nano-particle enhanced radiotherapy with protons, megavoltage photons and kilovoltage photons: A Monte Carlo simulation" by Lin et al [Phys. Med. Biol. 59 (2014) 7675–7689]"

Manuscript reference: PMB-114492

and

"Comment on "Biological modeling of gold nanoparticle enhanced radiotherapy for proton therapy" by Lin et al. [Phys. Med. Biol. 60 (2015) 4149–4168]"

Manuscript reference: PMB-114487

Thank you for your latest message regarding the two Comment articles you submitted.

Two Editorial board members have independently recommended that we do not consider your Comment articles. After further consultation with the Editorial board I can confirm that we will not be able to reconsider that decision.

6